\documentstyle[preprint,prl,aps,epsf]{revtex}

\begin{document}
 
\title{
Polymers in Curved Boxes
                                                }
\author{ K.\ Yaman$^*$ \and P.\ Pincus \\}
\address{
Department of Physics, University of California \\
Santa Barbara, CA 93106--9530 }
 
\author{ F.\ Solis \and T.A.\ Witten \\}
\address{
James Franck Institute, University of Chicago \\
5640 Ellis Avenue, Chicago, IL 60637}

\date{\today}
\maketitle
\begin{abstract}

We apply results derived in other contexts for the spectrum of the
Laplace operator in curved geometries to the study of an ideal polymer
chain confined to a
spherical annulus in arbitrary space dimension D and 
conclude that the free energy
compared to its value for an uncurved box of the same
thickness and volume, is
lower when $D < 3$, stays the same when $D = 3$, and is higher 
when \mbox{$D > 3$}. Thus confining an ideal polymer chain to a 
cylindrical shell,
lowers the effective bending elasticity of the walls, and
might induce spontaneous symmetry breaking, i.e. bending.
(Actually, the above mentioned results show that 
{\em {any}} shell in $D = 3$ induces this effect, 
except for a spherical shell). 
We compute the contribution of this effect to the bending rigidities in the 
Helfrich free energy expression.

\end{abstract}


\pagebreak

\section{Introduction}

\narrowtext

The quantum mechanics of particles confined to surfaces or surface
layers embedded in a higher dimensional space, e.g. curves, tubular
structures around curves, surfaces or layers between two surfaces in
three dimensions, has attracted attention from many authors in the
past three decades. 
Recently there has been considerable progress in rigorously, and
non-perturbatively analyzing the
changes to the spectrum of the Laplacian in curved geometries in two
dimensions \-- these have been motivated by applications in mesoscopic
semi-conductor physics, \cite{duclos}, \cite{renger}.
There are also studies of the
contribution of torsion to the bound state energies,
\cite{takagi}, \cite{clark}.
Until now the results for higher dimensional confinement are 
restricted to the perturbative regime,
\cite{homma}, \cite{jensen}, \cite{costa}, \cite{ogawa1}, \cite{ogawa2},
\cite{takagi}, \cite{tolar}.   

Below we summarize the main 
results of these studies, as they relate to the application of interest here. 
The next section reviews the relation between quantum mechanics and
polymer physics in some detail. We compute the free energy in the geometries
of interest in the third section. Most of the details of these calculations
are left to an appendix.  
The final section computes the effective 
bending rigidities of a dilute polymer solution confined in a box with elastic
walls, and includes our concluding remarks.

The correct quantization procedure for quantum particles confined to
manifolds embedded in higher dimensional spaces 
turns out to be non-trivial, \cite{homma}.
Studies, \cite{jensen} - \cite{exner}, which are perturbative in the 
curvature of the manifold,  conclude that 
confinement to
a curved manifold results in an effective potential,
which depends {\em both} on the manifold to which
the particle is confined {\em and} the particular way this manifold 
is embedded
in the higher dimensional space, i.e. {\em both} on the intrinsic 
{\em and} extrinsic curvatures of the manifold. 
This dependence on the extrinsic curvature means that one cannot arrive at
the correct Schr\"{o}dinger's equation by just
transforming the Laplacian to the coordinates of the confining
manifold; 
this can be understood in the light of the uncertainty
principle \-- the particle will know that there is an ``outside'', no
matter how strongly it is confined.

Confinement to a curve in three dimensions generates a
potential, which is attractive to the most highly curved regions,
in the one dimensional Schr\"{o}dinger's equation; this
potential is proportional to the square of the local
curvature of the curve, \cite{marcus}.  If one considers a `thin-tube'
that is wound around a curve and has constant cross-section, (see Fig.1),
to lowest order in the curvature one finds the
same result: 
the energy change is proportional to the square of
the curvature of the curve that the tube follows. 
The proportionality constant is independent of
the shape of the cross-section, and also of the exact details
of the confining potential (hard-walls etc.), 
\cite{costa}, \cite{takagi}, and \cite{gold} \-- this can be easily
understood by dimensional analysis, scaling arguments and 
using the result of \cite{marcus}.
Confinement between two
{\em {equidistant}} surfaces in general dimension results in a potential
$U$, which (in the appropriate units, i.e. neglecting overall positive
constants) is 
given to second order in the principal radii of curvatures $R_i$,
(\mbox{$i = 1,2,3,..,(D-1)$}), by \cite{jensen}: 

\begin{equation}
U = -\frac{1}{4} \,
\left ( 2 \sum_i \frac {1} {R^2_i} - 
{\left ( \sum_i \frac {1} {R_i} \right )}^2 \right ).
\label{potential}
\end{equation}
For a spherical shell embedded in $D$ dimensions one sets \mbox{$R_i = a$}, 
for all $i$, where $a$ is the radius of the shell, and, as 
first written down in \cite{tolar}, Eq.(\ref{potential}) reduces to
\mbox{$U = \frac{(D-1) (D-3)}{a^2}$}.  This is negative for $1 < D < 3$.

Of course, all of the above results apply to any physical system that
contains the Laplacian in its equations of motion.

\section{Connection to Polymer Physics}
 
The particle-in-a-box problem of quantum mechanics is closely related
to confined ideal polymers. A flexible polymer in solution is a random
walk. In the case without
excluded volume interactions this will be denoted as an `ideal chain';
i.e. ideal chains 
are allowed to self-intersect without raising their energy.
An ideal polymer chain configuration may be mapped onto the motion of 
a quantum
particle, where the probability distribution function is the analogue
of the wave function, and the degree of polymerization (N) plays the role of
imaginary time\cite{edwards}. 
In this spirit we reduce the `ideal-chain-in-a-box' problem to a 
quantum `particle-in-a-box' problem. 

We consider a random walk in space, and work in the
continuum description. Let $G(\vec{r}, \vec{r}\,', N)$ 
be the probability for a walk of $N$ steps to have its ends at $\vec{r}$ and
$\vec{r}\,'$. 
The function $G$ is defined to be $0$ for $N < 0$, and it  
satisfies \cite{edwards} the differential equation:

\begin{equation}
\left( {\partial \over {\partial N}} - 
{l^2 \over 6} \nabla^2_{\vec{r}} \right) G(\vec{r}, \vec{r}\,',N) =
\delta(\vec{r} -\vec{r}\,') \delta(N).
\label{diffuse}
\end{equation}
Here $l$ is the monomer length.
Eq.~(\ref{diffuse}) describes diffusion once the
boundary conditions are specified. We may also think about it
as Schr\"{o}dinger's equation, where $N$ plays the role 
of imaginary time. 

The partition function is just a sum over all configurations:

\begin{equation} 
Z = (1/V) \int d\vec{r} d\vec{r}\,' G(\vec{r}, \vec{r}\,', N).
\label{partition1}
\end{equation}
The factor of $1/V$ keeps the volume fixed when 
the free energy is computed.

Here we comment on the exact problem that we are considering, and
our choice of boundary conditions, when we compare
the flat and curved boxes. Our main aim is to find out the effects of
polymer confinement on the bending elasticities of the confining
structure. Hence we would like to isolate the change in the free
energy due to curvature {\em only}. Therefore we pick periodic
boundary conditions in the
transverse directions of the flat box. Also to the same end we conserve the
volume of the box during bending; this, of course, will fix the radius
of the curved box once
the amount of material (the surface area of the flat box) is fixed.
Assuming that the free energy per unit area for this system can be
expanded in the curvature when this radius is large compared to the
thickness of the box, one can compute the coefficients in this
expansion to second order in the curvature.
There are in general two bending elasticities to compute. 
These can be uniquely determined from the two
equations one attains by considering the free energy 
of the sphere and the cylinder.
We have in mind membranes which generally do not have free edges but end
at the walls
of a container, or are closed vesicles.
Then, e.g. for a container with repulsive walls, one
has to take into account when computing the energy balance the fact
that the eigenfunctions for the flat box vanish at the
transverse edges. 
Also, the volume for the real system may {\em{not}}
be conserved \-- this would bring translational entropy pieces into the
energy balance in addition to the contributions we calculate in the present
work.

To compute the partition function we use the method of eigenfunctions and
levels of the Schr\"{o}dinger's equation. This is an `effective Hamiltonian'
method where one integrates out certain configurations of the polymer chain
and labels the coarse-grained modes by discrete `levels'.
Thus we write 
$Z = {\displaystyle{\sum_n}} e^{\textstyle{- \frac {\epsilon_n} {k_B T}}}$, 
where the sum is over the levels 
and $\epsilon_n\,'s$ are the eigenvalues
of the effective Hamiltonian that describes the system in terms of
these levels only. 
For the remainder of this paper, in order to simplify the notation,
we measure all energies
in the units of $N k_B T$, and all lengths in the units of 
$(l/\sqrt{6})$, (except in the final results). 
It becomes clear what these levels are once one employs 
the eigenfunction expansion \cite{gennes} for the propagator of the chain: 
\mbox{$G(\vec{r}, \vec{r}\,', N) = {\displaystyle{\sum_n}} 
\Psi_{n}^{*}(\vec{r}) \Psi_{n}(\vec{r}\,') e^{- N E_n} \Theta(N)$}. 
Here $\Theta(N)$ is the step
function assuring that $G = 0$ for $N < 0$, and the 
$\Psi_{n}(\vec{r})$'s are normalized solutions to  
\mbox{$\nabla^2 \Psi (\vec{r}) = -E \Psi(\vec{r})$}. 
It is easy to verify that this
expansion indeed solves Eq.(\ref{diffuse}), 
(recall: \mbox{${\partial \over {\partial N}} \Theta(N) = \delta(N)$)}.
The boundary conditions are 
\mbox{$\Psi (\vec{r}) = 0$} at the walls which are at $|\vec{r}| = a$ and 
$|\vec{r}| = a + d$. This is a slab of thickness $d$ in one dimension, and a
spherical shell of radius $a$ and thickness $d$ in any dimension $D$.

Using this eigenfunction expansion in the partition function expression,
Eq.(\ref{partition1}), we find:  

\begin{equation}
Z =  (1/V) \sum_n e^{- N E_n} \int d\vec{r} \,\Psi^*_{n}(\vec{r})
                                \int d\vec{r\,'} \,\Psi_{n}(\vec{r\,'})=
(1/V) \sum_n e^{- N E_n} {\left| \int d\vec{r} \,\Psi_{n}(\vec{r})
                                                        \right|}^2.
\label{part}
\end{equation}
We see that in the geometry of interest all states except the spherically
symmetric ones 
disappear from the sum in Eq.~(\ref{part}), even though they do
contribute to the propagator. (In $D = 1$, the sinusoidal solutions with
an even number of peaks disappear). Hence:

\begin{equation}
Z = (1/V) \sum_{n=1}^{\infty} e^{- N E_n(a,d)} 
        {\left| \int dr \,r^{(D-1)} R_{n}(r) \right|}^2,
\label{partition}
\end{equation}
where $n$ labels the distinct s-wave levels, 
$R_{n}(r)$ is the normalized radial wavefunction, and we explicitly
indicate the dependence of the energies on $a$ and $d$. 
Defining:

\begin{equation}
B_n(a,d) \equiv { {\left| \int dr \,r^{(D-1)} R_{n}(r) \right|}^2 \over
{S_{D-1} \over D} ((a+d)^D - a^D)}  
\label{extra}
\end{equation}
where $S_{D-1}$ is the surface area of the unit sphere in $D$ dimensions, we
can write for the free energy:

\begin{equation}
F = - k_B T M \log\left(\,\sum^\infty_{n=1} 
                e^{-(E_n(a,d) N l^2/6)} B_n(a,d)\right),
\label{free}
\end{equation}
where M is the number of chains (each containing N links) in the box.

We now proceed with the analysis of the above expression.

\section{Free energy for polymers in boxes}

We consider a spherical annulus of large
radius and small thickness in arbitrary dimension D. The walls are
taken to be repulsive for the polymer. Physical interest
lies in $D = 2$ (Fig.~2), 
which is a cylindrical shell in three dimensions\footnote{The length along
the axis of the cylinder is taken to be large compared to the other
dimensions, and therefore does not play an essential role here.}, 
and $D = 3$ (Fig.~3), a spherical shell in three dimensions. 

The Laplacian in D-dimensions is $\nabla^2 = {d^2 \over dr^2} + 
{(D - 1) \over{r}} {d \over dr} - {\vec{L}^2 \over r^2}$. Here $\vec{L}$ is
the angular momentum operator in $D$ dimensions.
After standard manipulations we find the following quantization 
condition in terms of dimensionless variables: 

\begin{equation} 
J_\xi\left[({\tilde{E}})^{\frac{1}{2}} \, \frac{1}{\epsilon}\right] 
N_\xi \left[({\tilde{E}})^{\frac{1}{2}} \, 
\left(1+\frac{1}{\epsilon}\right)\right] 
- J_\xi\left[({\tilde{E}})^{\frac{1}{2}} \, 
\left(1+\frac{1}{\epsilon}\right)\right] 
N_\xi \left[({\tilde{E}})^{\frac{1}{2}} \, 
\frac{1}{\epsilon}\right] = 0,
\label{eigen}
\end{equation}
where $\tilde{E} \equiv E d^2$, $\xi^2  \equiv \nu^2 + ({{D - 2} \over 2})^2$,
$\epsilon \equiv ({d \over a})$, and $\nu^2$ is an eigenvalue of
the\footnote{SO(D) is the rotation group in $D$ dimensions.} 
$SO(D)$ symmetric Casimir operator, and the 
functions $J$ and $N$ are the standard Bessel functions. 
As explained earlier, spherical
symmetry of the geometries considered causes all but the $\nu = 0$ states to
drop out of the partition function.     
Defining $\mu \equiv {4 {\xi}^2}$, (for $\nu=0$), 
$\mu =  {(D - 2)}^2$, and labelling these s-wave states by $n = 1, 2, ...$, 
one can now proceed with a perturbative analysis by expanding 
Eq.(\ref{eigen}) in powers of $\epsilon = {d \over a}$, 
which results in:
\begin{equation}
d^2 E_n(a,d) = {(n \pi)}^2 +  
         {{(D-1) (D-3)} \over {4}} \left[ 
                \left({{d \over a}} \right)^2 -
                \left({{d \over a}} \right)^3 +
                \left({{d \over a}} \right)^4 
                \left(1 - {3 \over {8 n^2 \pi^2}} \right) + \cdots
                                                                \right].
\label{epert}
\end{equation}
This is the same perturbative result cited above. We see that the
curvature lowers the spherically symmetric energy spectrum for $1 < D <3$, 
does not affect it when $D = 1$ (trivial), or $D = 3$, and increases it 
everywhere else.
Notice that the perturbation series stops for $D = 1$, 
and $D = 3$. Also notice that the perturbative effect is the same for all
modes up to fourth order, this simplifies treating the full 
free energy of the polymer, Eq.(\ref{free}). This treatment
an expression for $B_n(a,d)$, the computation of which is difficult 
in general, but we were able to find an exact expression in $D = 3$, 
and a perturbative one in $D = 2$. We now summarize our results.
Details are left to the appendix.

\subsection{Three Dimensions}

Defining $Z_0$ as the partition function
of the flat box, and $k \equiv (N l^2 \pi^2 / 6 d^2)$ we find the exact 
result for the partition function:

\begin{equation}
Z/Z_0 = 1 + g(\epsilon) (1 - 3 f(k))
\label{three}
\end{equation}
where the functions $f$ and $g$  are defined in the appendix. One can
expand the above expressions in two limits: when the typical polymer
dimension,
$R_g \equiv \sqrt{N} l$, is smaller or larger than the box width, $d$. 
The latter is the so-called ground state dominance regime (G.S.D.) \--
here $k > 1$; the
former we will denote by "the colloid limit" (C.L.), 
since in this limit the coils look like small balls \-- here $k < 1$. 
In G.S.D., the corrections due to $B_n$ are negligible, and we find that the
free energy does not have any corrections due to curvature. In C.L.
we find:\\
$\Delta F = k_B T M \frac{1}{12} (\frac{d}{a})^2 
        \left(  \frac{8}{\pi^{1/2}} \sqrt{\frac{N l^2}{6 d^2}} -
                \frac{4 (3 \pi - 8)}{\pi} \frac{N l^2}{6 d^2} -
                \frac{16 (3 \pi - 8)}{\pi^{3/2}} 
                (\frac{N l^2}{6 d^2})^{3/2} + \cdots
                                                        \right)$                

\subsection{Two Dimensions}

In this case we find (to lowest order in $\epsilon$):

\begin{equation}
Z/Z_0 = e^{\textstyle{+\frac{k \epsilon^2}{4 \pi^2}}} 
\left(1 - {{\epsilon^2} \over {16}}
\left(1 - f(k) - \frac{8}{\pi^2}\, h(k)\right) + \cdots \right)
\label{two}
\end{equation}
where $f$ is the same function as for three dimensions, and $h$ is defined
in the appendix. This, in G.S.D. leads to:\\
$\Delta F = - k_B T M N \frac{1}{24} \frac{l^2}{a^2}$.\\
In C.L. there are no terms of order $\sqrt{k}$, and two terms of
order $k$ which cancel each other exactly, and the expansion starts at third
order, (this is similar to what happens for small hard spheres, i.e.
colloids, \-- there one
finds that there is no shift in the free energy in two dimensions, and a
positive shift in three dimensions, \cite{yaman}):\\
$\Delta F = - k_B T M \frac{1}{18\, \sqrt{6 \pi}} (\frac{d}{a})^2 
                        (\frac{N l^2}{d^2})^{3/2} + \cdots $\\
In the strong confinement (G.S.D.) regime it does not matter to which chain 
the monomers belong \-- the shift in the energy is proportional to the total
number of monomers, $M N$; in the C.L. regime we have an ideal gas of
$M$ objects.
 
We comment on some issues before computing effective bending rigidities 
using the above results: In the non-perturbative regime we solved the 
eigenvalue equation, Eq.~(\ref{eigen}), numerically for the ground states
in $D = 2, 4,$ and 5, for $2 < {a \over d} < 50$. The 
large ${a \over d}$ behavior of the solutions, (see the curves plotted
in Fig.~4), are 
exactly\footnote {This is not obvious from the plot given here.}
as predicted by
Eq.~(\ref{epert}). Notice that they are all converging to the flat case 
value of ${\pi}^2$ for very small curvature. As the curvature increases the 
behavior changes its functional form, and becomes much sharper than just a
quadratic deviation.
One can follow these curves farther towards infinite 
curvature, and see that they tend to a constant value that depends on the 
dimensionality. This pattern is also valid for $n \not= 1$, $\nu = 0$
(s-wave), states. 

For surfaces in three dimensions Eq.(\ref{potential}) reduces to:
\mbox{$U = - \frac{1}{4} {(\frac{1}{R_1} - \frac{1}{R_2})}^2$}. 
The effect of this potential,
even after integration over the entire surface, is always attractive
except for in two cases where it has no effect at all,
the flat surface, and the sphere. Notice also that in
\mbox{$D = 2$} one always gets an attractive potential if the confining curves
are not perfectly flat.  Thus in two dimensions, \--(physically we think
about
generalized cylinders living in three dimensions), the free energy is
reduced
by any curvature, regardless of the shape of the boundaries.
Although we are not considering large curvatures here, it is
worthwhile to mention that the literature cited in the introduction proves
that in one and two dimensional structures, (tubes and strips),
the ground state energy of the Laplace operator decreases due to
curvature, and there is a bound state localized in the regions of curvature. 
This is a non-perturbative statement, and it makes very
weak assumptions about the detailed geometry. This is
important because it implies that in these systems, 
(at least in G.S.D.), the polymer would localize 
in regions where there is curvature. One explicit example of this is the
numerical solution of Schr\"{o}dinger's equation in oval shaped rings
\cite{switkes}.

\section{Effect on $\kappa$ and ${\overline{\kappa}}$}

We conclude by calculating the contribution of this effect to the effective
bending curvature elasticity of an ideal polymer confined to a fluid
bilayer membrane. Let $f_{c}$ be the curvature free
energy per area. Up to quadratic order in the
principal radii ${1 \over {R_{1}}}$ and ${1 \over {R_{2}}}$,
$f_{c}$ can be written in terms of the mean and Gaussian curvatures of
the surface. In terms of the principal radii, the mean curvature is 
\mbox{$H = {1 \over 2} ( {1 \over {R_{1}}} + {1 \over {R_{2}}})$}, and the
Gaussian curvature is 
\mbox{$K =  ({1 \over {R_{1}}})  ({1 \over {R_{2}}})$}.
A general form of $f_{c}$ to this order is given by 
\cite{safran}: 
\mbox{$f_{c} = 2 \kappa ({H - c_{0}})^2 + {\overline{\kappa}} K$}, where 
$\kappa$ and ${\overline{\kappa}}$ are constants usually denoted by `bending
rigidities'. The spontaneous curvature of the surface, $c_{0}$, is taken as
zero for the rest of this discussion, because the film is assumed
symmetric with respect to its mid-plane. 

To find out the effects of
polymer confinement on $\kappa$, and
${\overline{\kappa}}$, we first consider a
vesicle (a spherical shell in  $(D = 3)$),  with a small curvature, i.e. a
shell of radius $a$, with a finite thickness $d$, where
$d \ll a$. For a sphere both principal radii are equal to $a$, and we
find 
\mbox{$f_c = {1 \over 2} \kappa ({2 \over a})^2 + {\overline{\kappa}}
({1 \over a})^2$}. 
Next consider a
cylindrical shell $(D = 2)$; now $R_{1} = a$, but $R_{2} = \infty$.
Thus, \mbox{$f_c = {1 \over 2} {\kappa \over a^2}$}.
Defining \mbox{$\Phi_M = M/V = M/(S d) + \rm{O}(\epsilon)$}, 
$S$ being the surface
area, and using the results for $\Delta F$ for the sphere and the cylinder
we conclude that:

\begin{equation}
\delta \kappa  = 
(k_B T) (\Phi_M d^3) \, \left(\frac{1}{8} \,
\left(1- f(k) - \frac{8}{\pi^2} h(k)\right) - \frac{k}{2 \pi^2} \right), 
\label{kap}
\end{equation}

\begin{equation}
\delta {\overline\kappa}=
(k_B T) (\Phi_M d^3) \left(-\frac{1}{6} + \frac{2}{\pi^2} h(k)\right),
\label{kapbar}
\end{equation}
where $\delta \kappa$ and $\delta {\overline{\kappa}}$ are the changes
in the bending rigidities. These expressions can be expanded in G.S.D.
and C.L., to give:

\begin{equation}
\delta \kappa  =
 - (k_B T) (\Phi_M l^3) \left\{ \begin{array}{lcl}
                N \,\frac{1}{12}\, (\frac{d}{l}) & ; & k > 1 \\
                N^{\frac{3}{2}} \, \frac{1}{9\, \sqrt{6 \pi}}
                                +\cdots & ; & k < 1
                                \end{array}
                                                \right.
\label{kapexp}
\end{equation}

\begin{equation}
\delta {\overline\kappa} =
 + (k_B T) (\Phi_M l^3) \left\{ \begin{array}{lcl}
                N \,\frac{1}{6}\, (\frac{d}{l}) & ; & k > 1 \\
                \sqrt{N} \,\frac{1}{3}\, \sqrt{\frac{2}{3 \pi}}
                        (\frac{d}{l})^2 -
                N \, \frac{3 \pi - 8}{18 \pi}\, (\frac{d}{l}) -
                N^{\frac{3}{2}} \, \frac{1}{9}\, ((3 \pi - 8) 
                        \sqrt{\frac{2}{3 \pi^3}} - 
                         \sqrt{\frac{2}{3 \pi}}) + \cdots
                                                                & ; & k < 1
                                \end{array}
                                                \right.
\label{kapbarexp}
\end{equation}

We have not considered any interactions between monomers. 
We expect that interactions between chains in a dense solution
would screen out the effect of the walls on the concentration profile
over a very short distance and hence weaken this curvature
effect. We are currently studying this situation.  
But our results may hold qualitatively
for concentrations below the overlap
concentration, $\Phi_m^{*}$. To be consistent with our analysis we use
Gaussian exponents to estimate $\Phi_m^{*}$: In G.S.D.
we require $ N \Phi_m^{*} = \frac{N}{R_F^2 \, d} = \frac{1}{l^2 d}$.
Thus $\Phi_m^{*} = \frac{1}{N \,l^2 \,d}$. In C.L. the walks
are three dimensional and we require 
$N \Phi_m^{*} = \frac{N}{R_F^3} = \frac{1}{\sqrt{N} l^3}$. 
Thus $\Phi_m^{*} = \frac{1}{N^{(3/2)} l^3}$. This leads to:

\begin{equation}
\frac{\delta \kappa}{(k_B T)}  =
 - \left\{ \begin{array}{lcl}
                \frac{1}{12} & ; & k > 1 \\
                 \sim 0.026 & ; & k < 1
                                \end{array}
                                                \right.
\label{kapexp2}
\end{equation}

\begin{equation}
\frac{\delta {\overline\kappa}} {(k_B T)}  =
 + \left\{ \begin{array}{lcl}
                \frac{1}{6} & ; & k > 1 \\
                0.06\, \frac{1}{N}\, {(\frac{d}{l})}^2& ; & k < 1
                                \end{array}
                                                \right.
\label{kapbarexp2}
\end{equation} 

Edwards theory assumes slow variations of the propagator
\cite{gennes}, and is therefore valid over
large length scales compared to the lattice spacing, $l$, and moderately
large number of steps, $N$. (Aside: We also need ${d \ll a}$, in order to 
use our perturbative result). Notice that for $k < 1$ we have
$\frac{1}{N}\,{(\frac{d}{l})}^2 > 1$ \--  which can be achieved e.g. with
$N \sim 100$, and $d/l \sim 20$, then this term gives a factor of $4$.
Estimating the bare rigidities of the box to be of the order of $k_B T$, 
we see that the effect is about $1$ \%
for $\kappa$, and $5$ \% for ${\overline{\kappa}}$.

We emphasize that the major result of this article is
that polymer confinement might reduce $\kappa$, and thereby induce 
spontaneous curvature in the system, such as
transitions from lamellar to bicontinuous phases or tubules etc,
provided that the effect persists after the inclusion of
self-avoidance, i.e. monomer-monomer interactions, in the analysis. We
expect that this would be the case below the overlap threshold, but 
future work should address this issue in detail. We
note here that, since there are two length scales in the problem, one
cannot solve the self-avoiding walk problem by simple scaling
arguments.

After one understands the role of interactions, one should also
consider the case of a polymer with a finite 
persistence length. We expect that in the regime where the
radius of curvature is much larger than the persistence length the
result would approach the one for the totally flexible polymer case 
we treated in
the present article (modulo interactions). But when the two lengths
are comparable, the aversity to bending of rigid polymers should
modify the effect computed here.

Note added in proof:
As we were preparing to publish we received a preprint by E. Eisenriegler, A.
Hanke, and S. Dietrich,  
in which the authors also compute bending rigidities 
of systems with chains near a repulsive surface.

\begin{acknowledgments}
We would like to thank C. Marques for useful discussions on several issues. 
KY and PP were supported by the National Science 
Foundation under UCSB MRL, DMR-9123048, and the Department of Energy
under DOE DE-FGO3-87ER45288. TW and FS acknowledge support from the National
Science Foundation under contracts DMR 92-08527 and DMR DMR-9400379.
\end{acknowledgments}

\appendix

\section{Details of the Free Energy Computation:}

We employ Eq.~(\ref{extra}) in three dimensions, which involves integrals
that are trivial to
do, and we find for $B_n$ the following expression:

\begin{equation}
B_n(a,d) = \frac{8}{n^2 \pi^2} 
                        \left\{ \begin{array}{lcl}
                1 + g(\epsilon) & ; & n = odd \\
                -3 g(\epsilon)  & ; & n = even
                                \end{array}
                                                \right.
\end{equation}
where 
$g(\epsilon) = - \frac{1}{4} \, \frac{\epsilon^3}{(1+\epsilon)^3 - 1}$.
Using this in Eq.~(\ref{part}) results in Eq.~(\ref{three}), where $f$ is
defined as follows:

\begin{equation}
f(k) \equiv 
\frac{\sum^{(e)}\frac{e^{- k n^2}}{n^2}}
        {\sum^{(o)} \frac{e^{- k n^2}}{n^2}}
\label{fthree}
\end{equation}
where the symbol $\sum^{(e)}$ denotes a sum over all positive even $n$, ($n
= 2, 4, 6,$ ...); and $\sum^{(o)}$ is over all positive odd $n$, ($n = 1, 3,
5, $...). 
As emphasized in the text Eq.~(\ref{three}) is exact with the function $f$
defined as above. This function can be expanded for large arguments (g.s.d.
regime) as:

\begin{equation} 
f(k) = \frac{1}{4} e^{-3 k}  ( 1 + \rm{O}(e^{- 8 k}) )
\end{equation}
For small
arguments one can also expand $f$ with a little more work. First we define
the function $A(t) \equiv 
\displaystyle{\sum_{n = -\infty}^{\infty}} e^{-t n^2}$.
This function has the following property:

\begin{equation} 
A(t) = \sqrt{\frac{\pi}{t}} A(\frac{\pi^2}{t})
\label{aidentity}
\end{equation}
Now we define $E(k) \equiv \sum^{(e)} \frac{e^{- k n^2}}{n^2}$, and 
$O(k) \equiv \sum^{(o)} \frac{e^{- k n^2}}{n^2}$. Then $f(k) =
\frac{E(k)}{O(k)}$. (Aside: $ \frac{8}{\pi^2}\, O(k)$ is the
partition function for the flat strip).
One can easily verify that $E(0) = \frac{\pi^2}{24}$,
and $O(0) = \frac{\pi^2}{8}$. Also,  
$\frac{\partial E(k)}{\partial k} = -\frac{1}{2} \, (A(4k) - 1)$, and 
$\frac{\partial O(k)}{\partial k} = -\frac{1}{2} \, (A(k) - A(4k))$.
Using Eq.(\ref{aidentity}), and the fact that 
$A(t) = 1 + \rm{O}(e^{-t})$, when $k$ is large, one finds that 
$\int_{0}^{k} \, dt A(t) = 2 \int_{0}^{k} \, dt A(4t) = 
2 \sqrt{\pi k} + \rm{O}(e^{-\frac{\pi^2}{k}})$, when 
$k$ is small. This allows an expansion of $f$:

\begin{equation}
f(k) = \frac{1}{3} \, 
\frac{1 - 12\,\pi^{-2}\, (\sqrt{\pi k} - k)}{1 - 4\, \pi^{-2}\, 
\sqrt{\pi k}} + \rm{O}(e^{-\frac{1}{k}}).
\label{fexp}
\end{equation}
This function crosses over to $\frac{1}{4} \, e^{-3 k}$ at $k \sim 0.3$.

Notice that these expressions satisfy the consistency checks:\\
$\epsilon = 0 \Rightarrow B_n(d) = \frac{8}{n^2 \pi^2}$, if $n$ is odd, and
$B_n = 0$, if $n$ is even. Also, $k = 0 \Rightarrow \Delta F = 0$. 

In two dimensions the integrals are also exactly doable but the final 
expression for $B_n$ is now much more complicated \-- we do not bother
including this here. 
It can, however, be expanded in $\epsilon$, to
yield: 

\begin{equation}
B_n(a,d) = \frac{8}{n^2 \pi^2}
                        \left\{ \begin{array}{lcl}
                1 - \frac{1}{16} \, \epsilon^2 
                + \frac{1}{2 \pi^2 n^2} \, \epsilon^2 +\cdots& ; & n = odd \\
                \frac{1}{16} \, \epsilon^2 + \cdots & ; & n = even
                                \end{array}
                                                \right.
\end{equation}

Now, when one is not in G.S.D., one in principle needs to worry about the 
terms that depend on $n$ in the expansion of the energy 
spectrum, Eq.(\ref{epert}). But when the argument of the terms in question
is larger than one, we are deep into the G.S.D. regime, whereas when it is
smaller than one these terms can be expanded in $\epsilon$, giving only
$\epsilon^4$ corrections. Thus, they can in practice be ignored when
computing bending rigidities. Defining:

\begin{equation}
h(k) \equiv \frac{\sum^{(o)} \frac{e^{- k n^2}}{n^4} }
                {\sum^{(o)} \frac{e^{- k n^2}}{n^2}}
\label{h}
\end{equation}
we find for the partition function the result in Eq.~(\ref{two}). Noticing
that $ - \frac{\partial h(k)}{\partial k} = O(k)$, and 
$h(0) = \frac{\pi^2}{12}$, one can expand $h$ using the expressions derived
above:

\begin{equation}
h(k) = \frac{\pi^2}{12}\, \frac{1 - 12 \, \pi^{-2}\, k +
32\, \pi^{-\frac{7}{2}} \, k^{\frac{3}{2}}}
{1 - 4 \pi^{-\frac{3}{2}}  \sqrt{k}} + \rm{O}(e^{-\frac{1}{k}}).
\label{hexp}
\end{equation}

Again, these expressions satisfy the above-mentioned consistency checks.
These expansions for $f$ and $h$ are in excellent agreement with numerical
results for $0 < k < \frac{1}{\pi}$.
The results for the bending rigidities follow from these expansions.

\pagebreak

{\bf {FIGURE CAPTIONS}}\\

\hspace*{- \parindent}
Fig.1 : Twisting tube example: solid cylinder $\longleftrightarrow$ torus.\\ 
Fig.2 : $D = 2$ annulus. \\
Fig.3 : $D = 3$ annulus. \\
Fig.4 : Numerical solution for $\tilde{E}(1/\epsilon)$, in two, four, and
five dimensions.\\
\epsfverbosetrue
\pagebreak
\vspace{10pt}
\epsfxsize = 3in
\centerline{ \epsfbox{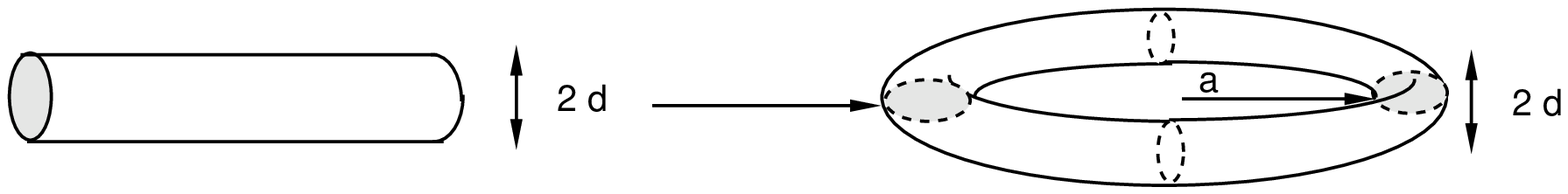}}
\vspace{12pt}

\pagebreak
\vspace{10pt}
\epsfxsize = 3in
\centerline{ \epsfbox{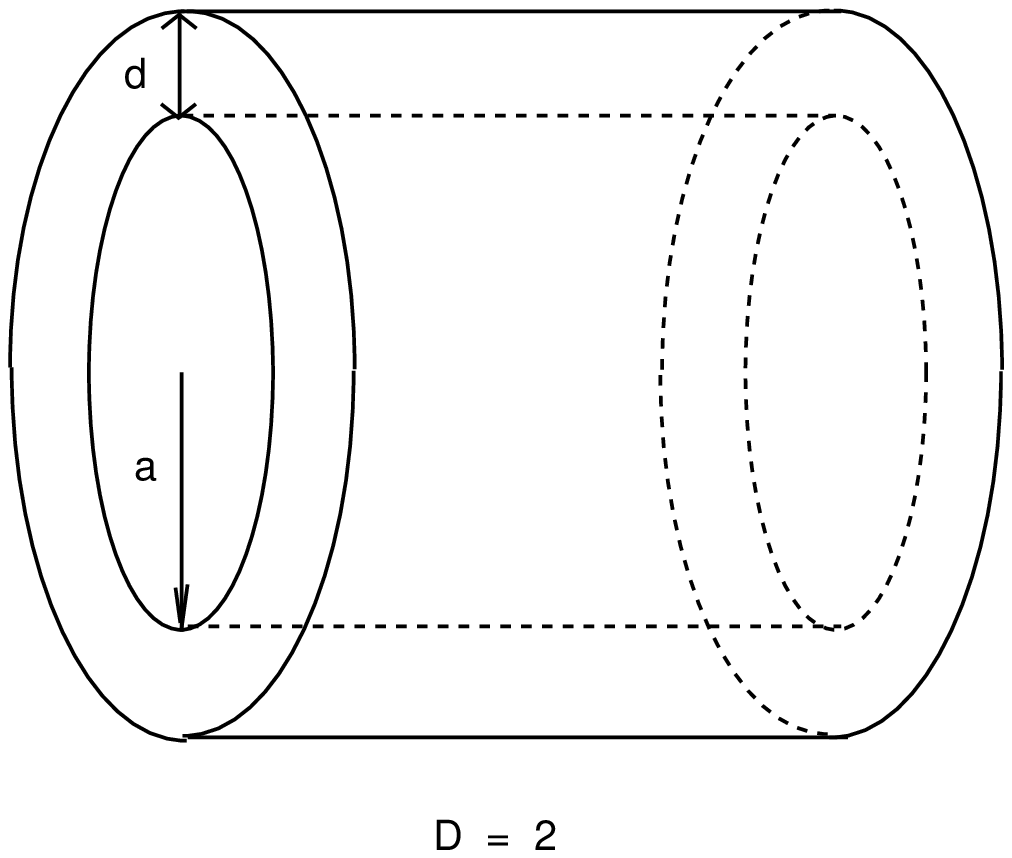}}
\vspace{12pt}

\pagebreak
\vspace{10pt}
\epsfxsize = 3in
\centerline{ \epsfbox{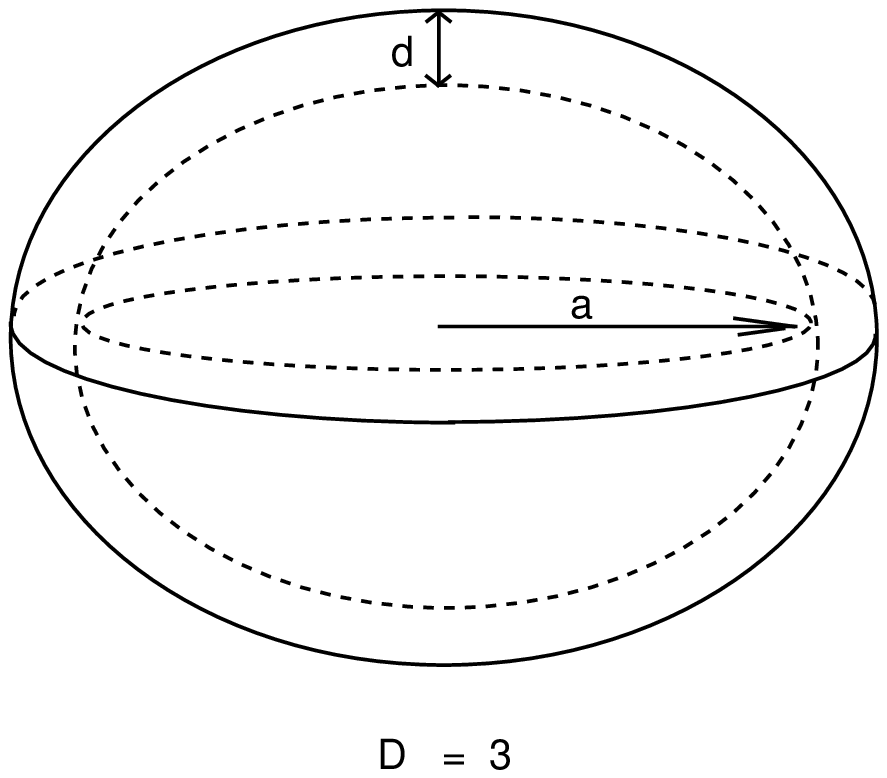}}
\vspace{12pt}

\pagebreak
\vspace{10pt}
\epsfxsize = 3in
\centerline{ \epsfbox{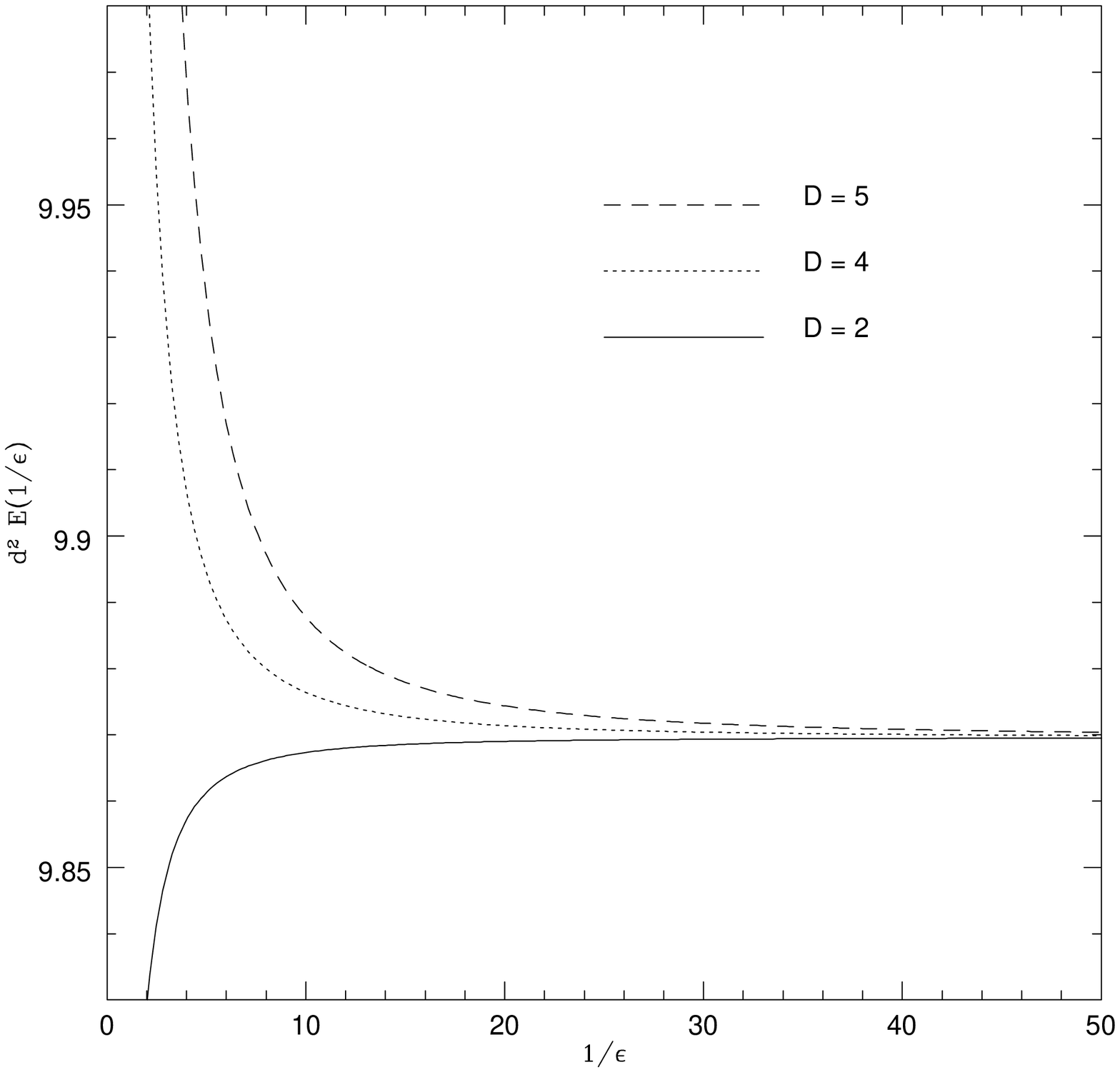}}
\vspace{12pt}


\begin{thebibliography}{99}

\bibitem{duclos}Duclos, P.; Exner, P. \,{\em{Rev. Math. Phys.}} 
${\bf{1995}}, {\it{7}}, 73$.

\bibitem{renger}Renger, W.; Bulla, W. \,{\em{Lett. Math. Phys.}} 
${\bf{1995}}, {\it{35}}, 1$.

\bibitem{takagi}Takagi, S.; Tanzawa, T. \,{\em{Prog. Theor. Phys.}}
${\bf{1992}}, {\it{87}}, 561$.

\bibitem{clark}Clark, I.J.; Bracken, A.J. \,{\em{J. Phys. A}} 
${\bf{1996}}, {\it{29}}, 339$.

\bibitem{homma}Homma, T.; Innamoto, T.; Miyazaki, T. \,{\em{Phys. Rev. D}}
${\bf{1990}}, {\it{42}}, 2049$.

\bibitem{jensen}Jensen H.; Koppe, H. \,{\em{Ann. of Phys.}}
${\bf{1971}}, {\it{63}}, 586$.

\bibitem{costa}daCosta, R.C.T. \,{\em{Phys. Rev. A}}
${\bf{1981}}, {\it{23}}, 1982$.
                                                           
\bibitem{ogawa1}Ogawa, N.; Fujii, K.; Kobushukin A.\, 
{\em{Prog. Theor. Phys.}}
${\bf{1990}}, {\it{83}}, 894$.

\bibitem{ogawa2}Ogawa, N.; Fujii, K.; Chepilko, N.; Kobushukin, A.\,
{\em{Prog. Theor. Phys.}} 
${\bf{1991}}, {\it{85}}, 1189$.

\bibitem{tolar}Tolar, J. \, 
{\em{On a quantum mechanical d'Alembertian principle}},
in {\em{Group Theoretical Methods in Physics}}, Lecture Notes in Phys.
${\it{313}}$, (Springer Verlag, Berlin, ${\bf{1988}}$), pp.$268-274$.

\bibitem{marcus}Marcus, J.\, {\em{J. Chem. Phys.}} 
${\bf{1966}}, {\it{45}}, 4493$.

\bibitem{exner}Exner, P.; Seba, P.; Stovicek, P.\, {\em{Phys. Lett. A}}
${\bf{1990}}, {\it{150}}, 179$.

\bibitem{gold}Goldstone, J.; Jaffe, R.L.\, {\em{Phys. Rev. B}}, 
${\bf{1992}}, {\it{45}}, 14100$.

\bibitem{switkes}Switkes, E.; Russell, E.L.; Skinner, J.L.\, 
{\em{J. Chem. Phys.}}
${\bf{1977}}, {\it{67}}, 3061$. 

\bibitem{edwards}Edwards S.F.; Doi, M.\, 
{\em{The Theory of Polymer Dynamics}} 
(Oxford Science Publications, New York, NY, 
${\bf{1986}}$), p.$19$.

\bibitem{gennes}deGennes, P.G.\,
{\em{Scaling Concepts in Polymer Physics}} 
(Cornell University Press, Ithaca, NY, ${\bf{1979}}$). 

\bibitem{yaman}Yaman, K.; Marques, C. \,Unpublished results.

\bibitem{safran}Safran, S.A.\, 
{\em{Statistical Thermodynamics of Surfaces, Interfaces, and Membranes}} 
(Addison-Wesley Publishing Co., ${\bf{1994}}$), p.$189$.
 
\end{thebibliography}
\end{document}